\documentclass[final,5p,times,twocolumn]{elsarticle}
\pdfoutput=1
\usepackage{placeins}
\usepackage{graphicx}
\usepackage{verbatim}
\usepackage{epsfig}
\usepackage{subfigure}
\usepackage{color}
\usepackage{multirow}
\usepackage{comment}

\usepackage{slashed}
\usepackage{amsmath}
\usepackage[normalem]{ulem}
\usepackage{tikz}

\usepackage{cuted}
\usepackage{amssymb,amsmath,amsfonts,mathrsfs}

\usepackage{longtable}
\usepackage{verbatim}
\usepackage{color}
\usepackage[mathscr]{euscript}
\usepackage{framed}
\usepackage{mdframed}  

\usepackage{cancel}
\usepackage{color}
\usepackage{hyperref}
\hypersetup{colorlinks, citecolor=bluscuro, linkcolor=black, urlcolor=bluscuro}
\definecolor{rossos}{cmyk}{0,1,1,0.55}
\definecolor{bluscuro}{rgb}{0.15, 0.2, .85}
\definecolor{bluchiaro}{cmyk}{1,.3,0.,0.1}
\graphicspath{{./Figures/}}
\allowdisplaybreaks

\numberwithin{equation}{section}
\renewcommand\theequation{\arabic{section}.\arabic{equation}}

\usepackage{tikz}

\usepackage{tcolorbox}

\journal{Nuclear Physics B}
\setcitestyle{open={[},close={]}}

\begin{document}

\begin{frontmatter}

\title{ \huge    Non-linear Black Hole Ringdowns:\\
 an Analytical Approach}

\author[a,b]{D. Perrone,}
\author[c]{T. Barreira,}
\author[d]{A. Kehagias,}
\author[a,b,e]{A. Riotto}

\ead{davide.perrone@unige.ch}
\ead{thomas.barreira@polytechnique.edu}
\ead{kehagias@central.ntua.gr}
\ead{antonio.riotto@unige.ch}

\affiliation[a]{
	Department of Theoretical Physics and Center for Astroparticle Physics (CAP) \\
			24 quai E. Ansermet, CH-1211 Geneva 4, Switzerland}
\affiliation[b]{Gravitational Wave Science Center (GWSC), Universit\'e de Gen\`eve, CH-1211 Geneva, Switzerland}

\affiliation[c]{\'Ecole Polytechnique,  91128 Palaiseau Cedex, France} 

\affiliation[d]{Physics Division, National Technical University of Athens, Athens, 15780, Greece}

\affiliation[e]{LAPTh, CNRS, USMB, F-74940 Annecy, 
France} 

\begin{abstract}
    Due to the  nature of gravity, non-linear effects are left imprinted in the quasi-normal modes generated in  the ringdown phase of the merger of   two black
holes.   We offer an analytical treatment of  the quasi-normal modes at second-order in black hole perturbation theory which takes advantage from the fact that the non-linear sources are peaked around the light ring. As a byproduct, we describe why  the  amplitude of the second-order mode relative to the square of the first-order amplitude depends only weakly on the initial condition of the problem.  
\end{abstract}

\vskip 2cm

\end{frontmatter}
\newpage
\section{Introduction}
During the  final stage,   called ringdown, of the merger of two Black Holes (BHs), Quasi Normal Modes (QNMs)    describe  the BH  response to any kind of disturbance and their frequencies only depends on the BH mass, spin and charge of the final BH\cite{Barack:2018yly}. As such, QNMs  provide  a privileged  tool  to understand  the properties of BHs, which is one  of the   major goals of gravitational wave astronomy \cite{Maggiore:2018sht}.

The  Gravitational Wave (GW)  strain   is generically  described using first-order BH perturbation theory. However, non-linearities are  an intrinsic property of general relativity. As a consequence,     second-order
effects turn out to be  relevant   in describing  ringdowns from BH mergers and have received much attention recently  \cite{Cheung:2022rbm,Mitman:2022qdl,Lagos:2022otp,Baibhav:2023clw,Kehagias:2023ctr,Kehagias:2023mcl,Khera:2023lnc}
(see  Refs. \cite{08article, Nicasio:1998aj,nl1, Nakano:2007cj,nl2, nl3, Berti_2009, Kokkotas_1999} for some older results). For example, from numerical results like those of \cite{Cheung:2022rbm, Mitman:2022qdl}, it can be seen that non-linearities could account for a relevant fraction of the amplitude signal, being approximately proportional to 10\% or more of the linear amplitude squared.

The GWs produced during the  ringdown and sufficiently far from the BH horizon are the  sum of exponentially damped QNMs, characterized  by  two angular harmonic numbers $(l,m)$ and an overtone number $n$. Their amplitude is denoted $A_{(l,m,n)}$ and  their oscillation frequency and decay timescale are given by  the real and imaginary parts of~$\omega_{(l,m,n)}$, respectively, 

\begin{eqnarray}
r\,h_{l,\,m}(t-r)&=&\sum_{n\geq 0}A_{l,m,n} e^{-i\omega_{l,m,n}(t-r)}.
\end{eqnarray}
The non-linearities are generically caught by the ratios

\begin{equation}
\label{ratio}
{\rm NL}_h =\left|\frac{A_{l_1,m_1,n_1\times l_2,m_2,n_2}}{A_{l_1,m_1,n_1}A_{l_2,m_2,n_2} }\right|, 
\end{equation}
where the amplitudes $A_{l_1,m_1,n_1\times l_2,m_2,n_2}$ of the second-order non-linear GW strain is constructed from the linear GWs with amplitudes $A_{l_1,m_1,n_1}$ and $A_{l_2,m_2,n_2}$ with frequency 

\begin{equation}
\omega_{{l_1,m_1,n_1\times l_2,m_2,n_2}}=\omega_{l_1,m_1,n_1}+\omega_{l_2,m_2,n_2}.
\end{equation}
Some generic  outputs have been pointed out in the literature, independently from the properties of the BH or from the type of the BH merger, be it originating from the quasi-circular merger of two BHs giving rise to a  Kerr BH or  from  head-on spinless BH collisions resulting in a spinless BH:

\begin{enumerate}
    \item 
the amplitude $A_{l_1,m_1,n_1\times l_2,m_2,n_2}$ scales like $A_{l_1,m_1,n_1}A_{l_2,m_2,n_2}$, such that in the ratio NL the amplitudes cancel out;  

\item the relative
amplitude of second-order modes depends mildly  on the initial conditions \cite{Cardoso_new}. 

\end{enumerate}

The goal of this paper is to offer an analytical treatment of the second-order QNMs and an explanation of these two general results. We will do so by offering some generic considerations about BH perturbation theory at second-order and showing that, upon mild assumptions and based on the nature of the non-linear sources, properties 1 and 2 can be recovered. We will also offer more quantitative results by using the WKB approximation.  

The paper is organized as follows. 
In section 2 we offer generic considerations and focus on the first-order solution. Section 3 contains the description of the second-order evolution. In section 4 we devote our attention on the dependence on the initial conditions, while the WKB approximation is used in section 5. Section 6 provides various examples, and section 7 contains our conclusions. Finally, the paper is supplemented by various appendices.
We will use units of $G=\hbar = 1$

\section{Setting the stage}
We start by writing the BH equations for the linear and quadratic perturbations, $\psi_1$ and $\chi_2$ respectively, in their generic and somewhat symbolic form \cite{Maggiore:2018sht}
\begin{equation}
\label{linear}
   ( \partial^2_x - \partial^2_t - V(x) )\,\psi_1(x,t) = 0,
\end{equation}
\begin{equation}
\label{nlinear}
   ( \partial^2_x - \partial^2_t - V(x) )\,\chi_2(x,t) = S(\psi_1^2),
\end{equation}
where the subscript indicates the order in perturbation theory and where the tortoise coordinate is defined as
\begin{equation}
    x = r + 2M \log\left(\frac{r}{2M}- 1\right).
\end{equation}
For practical purpouses we will use   in the following the  Zerilli potential $V(r(x))$
\begin{eqnarray}
    V(r) &=& \left(1-\frac{2M}{r}\right) \;\times \\
    &\times &\frac{2\lambda^2 (\lambda+1)r^3 + 6 \lambda^2 M r^2 + 18 \lambda M^2 r + 18 M^3}{r^3 (\lambda r + 3M)^2},\nonumber\\
    \lambda &=& \frac{1}{2}(l-1)(l+2),
\end{eqnarray}
where $l$ is the angular momentum number, but our considerations are not limited by this choice.
We have introduced  a generic source term $S$, which is a quadratic function of the first-order solution $\psi_1(x,t)$ and  its derivatives in time and in space.
To fully solve the problem we need to also specify two initial conditions for the first- and second-order solutions
\begin{equation}
    \psi_1(x,0) = f_1(x), \quad \dot{\psi}_1(x,0) = g_1(x),
\end{equation}
\begin{equation}
    \chi_2(x,0) = f_2(x), \quad \dot{\chi}_2(x,0) = g_2(x),
\end{equation}
and a set of boundary conditions, which will depend on the problem we would like to solve.
In our case we are interested in QNMs, and  we will take purely ingoing boundary conditions at the horizon $r=2M$ and purely outgoing at infinity $r\to \infty$. Schematically, using the tortoise coordinate and going in frequency space
\begin{equation}
    \psi(x,\omega) \to e^{i\omega x}, \quad x\to \infty,
\end{equation}
\begin{equation}
    \psi(x,\omega) \to e^{-i\omega x}, \quad x\to -\infty.
\end{equation}
We will now proceed by the Laplace technique whose main properties  we will briefly summarise for the reader's sake.

\subsection{Laplace transform}
We define a Laplace transform operator as
\begin{equation}
    \mathcal{L}[f] =\tilde{f}(s)= \int_0^{\infty}{\rm d} t \,e^{-st}\, f(t),
\end{equation}
and its inverse as
\begin{equation}
    \mathcal{L}^{-1}[\tilde{f}] =f(t)= \int_{\epsilon-i\infty}^{\epsilon + i\infty} {\rm d} s \, e^{st} \,\tilde{f}(s), 
\end{equation}
where the integral is performed in the complex s plane and $\epsilon\to 0^+$. Usually to perform the integral one closes  it with a semi-circular path sent to infinity, but one can also deform it to take into account the presence of branch cuts on the negative real  axis. 
Furthermore, one can think of the $s$ variable as $s\sim -i\omega$, to connect with the usual Fourier analysis. In this case the causality properties are linked to absence of poles in the half-plane $\text{Re}(s)>0$.
A very useful property of Laplace transform is
\begin{equation}
\label{laplprop1}
    \mathcal{L}[f'] = \int_0^{\infty} {\rm d} t \, e^{-st} \,\partial_t f(t)  = - f(0) + s  \mathcal{L}[f],
\end{equation}
which will allow us to introduce the initial conditions in the differential equation as an extra source term.

\subsection{The  first-order solution}
We start by focusing on the first-order problem. Most of the material in this section is standard and the expert reader can skip it.  

We first perform the Laplace transform in the variable  $t$, to get
\begin{equation}
   ( \partial^2_x - s^2 - V(x))\,\psi_1(x,s) = -s f_1(x) - g_1(x),
\end{equation}
where the initial conditions enter in the equation as a source term because of Eq. (\ref{laplprop1})  \cite{Szpak:2004sf}. We define the source term for the linear problem as
\begin{equation}
\mathcal{J}_1(x,s) = -s f_1(x) - g_1(x),
\end{equation}
to write the differential equation in the familiar form.
\begin{equation}
   ( \partial^2_x - s^2 - V(x))\,\psi_1(x,s) = \mathcal{J}_1(x,s). 
\end{equation}
It is also straightforward to write purely ingoing and outgoing boundary conditions in Laplace space,
\begin{equation}
    \psi_1(x,s) \to e^{- s x}, \quad x\to \infty,
\end{equation}
\begin{equation}
    \psi_1(x,s) \to e^{s x}, \quad x\to -\infty\;.
\end{equation}
We can now look for the corresponding Green's function 
\begin{equation}
   ( \partial^2_x - s^2 - V(x))\,G(x,x',s) = \delta(x-x').
\end{equation}
To find it in Laplace space we solve first the homogeneous problem
\begin{equation}
   ( \partial^2_x - s^2 - V(x))\,\phi_{\pm} = 0,
\end{equation}
where $\phi_+$ is the solution of the homogeneous equation with only the boundary condition at $x\to +\infty$ and $\phi_-$ is the one with the boundary condition at $x\to -\infty$. This means that the asymptotic behaviour of the solutions is
\begin{align*}
&\phi_+(x,s) \to e^{-sx}, \quad \quad \qquad \qquad \; \; x\to +\infty, \\
&\phi_+(x,s) \to A_+(s)\; e^{sx} + B_+(s) \;e^{-sx} ,\quad x\to -\infty,
\end{align*}
\begin{align*}
&\phi_-(x,s) \to A_-(s)\; e^{sx} + B_-(s) \;e^{-sx},\quad x\to +\infty, \\
&\phi_-(x,s) \to e^{sx}, \quad \quad \qquad \qquad \; \; x\to -\infty.
\end{align*}
The Green's function can be written as
\begin{equation}
G(x,x',s) = \frac{\phi_+(x_>,s)\, \phi_-(x_<, s)}{W(s)},
\end{equation}
where 
\begin{equation}
x_> = \max(x, x'), \quad x_< = \min(x, x'),
\end{equation}
and we have introduced the Wronskian of the solution, as a function of $s$

\begin{equation}
W(s) = \phi_-(x,s)\; \partial_x \, \phi_+(x, s)\; - \phi_+(x,s)\; \partial_x \, \phi_-(x, s).\\
\end{equation}
Notice that in this type of differential equations the Wronskian is independent of the specific $x$ chosen, and it will have a zero in $s$ whenever the two solutions $\phi_{\pm}$ coincide, satisfying both boundary conditions at the same time. We proceed by performing a convolution of the Green's function  with the source term, containing our initial conditions. This will give us the solution of the differential equation in Laplace space
\begin{equation}
\psi_1(x,s) = \int_{-\infty}^{+\infty} \; {\rm d}x'\; G(x, x', s)  \mathcal{J}_1(x',s),
\end{equation}
or, explicitly,
\begin{equation}
\psi_1(x,s) = \int_{-\infty}^{+\infty} \; {\rm d}x'\;  \frac{\phi_+(x_>,s)\, \phi_-(x_<, s)}{W(s)} \left( -s\,f_1(x') - g_1(x') \right).
\end{equation}
Using the inverse Laplace transform we have the solution at first-order
\begin{align}
\psi_1(x,t) &=  \int_{\epsilon - i \infty}^{\epsilon + i \infty} \; {\rm d} s\; e^{ts} \; \psi_1(x,s) =\nonumber \\ 
&= \int_{\epsilon - i \infty}^{\epsilon + i \infty} \; {\rm d} s\; e^{ts} \; \int_{-\infty}^{+\infty} \; {\rm d}x'\;  \frac{\phi_+(x_>,s) \phi_-(x_<, s)}{W(s)}\;\times \nonumber\\
&\times\;\left( -s\,f_1(x') - g_1(x') \right).
\end{align}
As we integrate in the complex $s$ plane we will pick up poles of the function $\psi_1(x,s)$ and eventual contour deformations if branch cuts are present. If we assume that the numerator has no poles in $s$, we are  left with the zeros of the Wronskian $W(s)$ in the denominator, which give us the damped oscillatory behaviour of the QNMs. The branch cuts that could arise from the numerator are linked to the subleading polynomial behaviour which is observed after the ringdown.
We will ignore the effects of such polynomial behaviours  because our goal is to confront the analytic calculation with a specific ratio of amplitudes extracted from the dominant oscillating behaviour in numerical experiments.

Picking poles of the Wronskian forces $\phi_+ = \phi_-$, and we can write
\begin{align}
\psi_1(x,t) =& \sum_{n} e^{ts_n} \; \frac{1}{W'(s_n)}\; \int_{-\infty}^{+\infty} \; {\rm d}x'\;  \phi_{-}(x,s_n)\, \phi_-(x', s_n)\;\times \nonumber\\
&\times \left( -s_n\,f_1(x') - g_1(x') \right).
\end{align}
and the distinction between $x_>$ and $x_<$ can be dropped, choosing one of the two $\phi$'s as function of $x$ and the other as function of $x'$.
If we are  interested only in the  exponential behaviour, we can define
\begin{equation}
c_{n{\rm L}}=W'(s_n)^{-1}\; \int_{-\infty}^{+\infty} \; {\rm d}x'\; \phi_{-}(x', s_n) \left( -s_n\,f_1(x') - g_1(x') \right),
\end{equation}
and build up the solution in a series expansion as
\begin{equation}
\psi_1(x,t) = \sum_{n} \;c_{n{\rm L}}\; \phi_{-}(x,s_n) \;e^{ts_n} .
\end{equation}
In the following sections we will further approximate this result by picking only the first zero of the Wronskian $s_0$, the fundamental mode, because it is the least damped  at infinity

\begin{equation}
\psi_1(x,t) \simeq c_{0{\rm L}}\; \phi_{-}(x,s_0) \;e^{ts_0}.
\end{equation}

\begin{figure}
    \centering
    \includegraphics{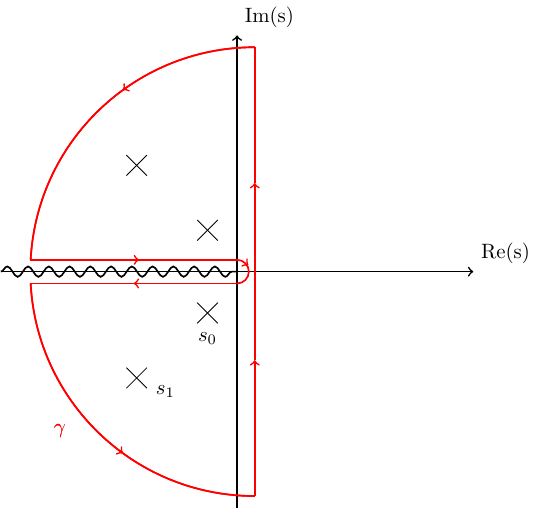}
    \caption{Typical pole structure of the solutions $\psi(x,s)$.}
    \label{fig:poles}
\end{figure}

\subsection{Source term}
 In full generality, the source is  the   sum of terms of the type
\begin{equation}
S(x,t) = \sum_{i=0}^{N_{\rm max}} R_i(r(x)) \left(D_{i,1}[t, x] \psi_1 (x,t) \right) \left(D_{i,2}[t, x] \psi_1 (x,t) \right),
\end{equation}
where $D[t,x]$ is a differential operator applied to the first-order solution, and $R_i(r)$ is a generic ratio of polynomial functions in the $r$ variable. The differential operators $D[t,x]$ can always be put in the form
\begin{equation}
D[t,x] = \partial_t \;  \partial_x^j \quad \text{or} \quad D[t,x] = \partial_x^k\,,
\end{equation}
with $(j,k)$ generic non-negative integers, because we can always exchange $\partial_t^2$ with $(\partial_x^2 - V(x))$, being $\psi_1$ a solution for the first-order differential equation.
Inserting the explicit form of $\psi_1(x,t)$ found previously and approximating it further by keeping only the dominant mode
\begin{equation}
\psi_1(x,t) \simeq c_{0{\rm L}}\; \phi_-(x,s_0) \;e^{ts_0},
\end{equation}
we get
\begin{equation}
\label{ss1}
S(x,t) = \left(c_{0{\rm L}} e^{s_0 t} \right)^2 \; 
h(x),
\end{equation}
where 
\begin{equation}
h(x)=\sum_{i=0}^{N_{\rm max}} R_i(r(x))\; (\partial_x^{k_i} \phi_- (x,s_0) )\; (\partial_x^{j_i} \phi_- (x,s_0)).
\end{equation}
It is  interesting to notice that in general we have a factorisation of the constant $c_{0{\rm L}}$ which has all the information about the initial conditions and we also have the expected oscillation with frequency $2 s_0 \sim 2\omega_{lm0}$. Furthermore we know that  the source is peaked  around a certain $x_0$, which is where the light-ring is placed and approximately at the peak of the potential $V(x)$.

\section{The evolution at second-order}
Having set the stage, we are  now ready to solve the second-order problem

\begin{equation}
   ( \partial^2_x - \partial^2_t - V(x) )\,\chi_2(x,t) = S(x,t).
\end{equation}
 We start again with the Laplace transform in $t$, which gives 
\begin{equation}
  ( \partial^2_x - s^2 - V(x) )\,\chi_2(x,s) = \mathcal{J}_2(x,s),
\end{equation}
where now the source $\mathcal{J}_2$ is
\begin{equation}
\mathcal{J}_2(x,s) = - s f_2(x) - g_2(x) + S(x,s)
\end{equation}
and the Laplace transform of the source is simply
\begin{equation}
S(x,s) = \frac{1}{s-2s_0} c_{0{\rm L}}^2 \; h(x).
\end{equation}
Since the Green's function is the same as at first-order,   the solution at second-order is just
\begin{equation}
\chi_2(x,s) = \int_{-\infty}^{+\infty} \; {\rm d}x'\; G(x, x', s) \, \mathcal{J}_2(x',s),
\end{equation}
which explicitly reads 
\begin{align}
\chi_2(x,s) = \int_{-\infty}^{+\infty} \; {\rm d}x'\;  \frac{\phi_+(x_>,s)\, \phi_-(x_<, s)}{W(s)} \times \nonumber \\
\times \;\left( -s\,f_2(x') - g_2(x') + S(x',s) \right).
\end{align}
Performing the inverse Laplace transform we can identify two types of contribution, depending on the pole structure
\begin{equation}
\chi_2(x,t) = \chi_2^I(x,t) + \chi_2^S(x,t),
\end{equation}
where the first is the term obtained from the initial conditions
\begin{align}
\chi_2^{I}(x,t) =\int_{\epsilon - i \infty}^{\epsilon + i \infty} \; {\rm d}s\; e^{ts}\; \int_{-\infty}^{+\infty} \; {\rm d}x'\;  \frac{\phi_+(x_>,s)\, \phi_-(x_<, s)}{W(s)} \times \nonumber \\
\times \;\left( -s\,f_2(x') - g_2(x') \right),
\end{align}
and the second comes from the source
\begin{equation}
\chi_2^S(x,t) =\int_{\epsilon - i \infty}^{\epsilon + i \infty} \; {\rm d}s\; e^{ts}\; \int_{-\infty}^{+\infty} \; {\rm d}x'\;  \frac{\phi_+(x_>,s)\, \phi_-(x_<, s)}{W(s)} S(x',s).
\end{equation}
Let us  look at  them separately,  starting with the contribution from initial conditions.
It has exactly the same analytic structure in $s$ as the one obtained at linear order, so it will contribute to the solution with frequency $s_0$, which will of course depend upon the angular momentum we are considering. Exactly as before we can integrate over the poles of $1/W$, sending $\phi_+ = \phi_-$, with the same analyticity assumptions on the functions we made for the linear case. Consequently, 
\begin{align}
\chi_2^I(x,t) = \sum_{n} e^{ts_n} \; \frac{\phi_-(x,s_n)}{W'(s_n)}\; \int_{-\infty}^{+\infty} \; {\rm d}x'\;   \phi_-(x', s_n)\times \nonumber\\
\times \;\left( -s_n\,f_2(x') - g_2(x') \right).
\end{align}
Defining 
\begin{equation}
c_{n{\rm NL}}=W'(s_n)^{-1}\; \int_{-\infty}^{+\infty} \; {\rm d}x'\; \phi_-(x', s_n) \left( -s_n\,f_2(x') - g_2(x') \right),
\end{equation}
we  get
\begin{equation}
\chi_2^I(x,t) = \sum_{n} \;c_{n{\rm NL}}\; \phi_-(x,s_n) \;e^{ts_n} .
\end{equation}
\\
As for the source term,  we have two contributions for the poles, the zeros of the Wronskian and the pole of the source. Separating them we get
\begin{align}
\chi_2^S(x,t)\Big|_{W \text{ poles}} =\sum_{n}  \frac{1}{W'(s_n)}\;e^{ts_n}\; \phi_-(x,s_n)\; \times \nonumber \\
\times \;\int_{-\infty}^{+\infty} \; {\rm d}x'\;  \phi_-(x', s_n) S(x',s_n),
\end{align}
and
\begin{align}
\chi_2^S(x,t)\Big|_{S \text{ pole}} = c_{0{\rm L}}^2 e^{2s_0 t} \; \int_{-\infty}^{+\infty} \; {\rm d}x'\;  \frac{\phi_+(x_>,2s_0)\, \phi_-(x_<, 2s_0)}{W(2s_0)} \; h(x').
\end{align}
At this stage, it is  crucial that the spectrum $s_n$ coming from the Wronskian have no pole corresponding to $2s_0$, otherwise we shall end up with a double pole. This situation does not happen for the QNM spectrum, at least for the dominant frequencies.

For the Wronskian poles part we can define
\begin{align}
c_W &= W'(s_n)^{-1} \; \int_{-\infty}^{+\infty} \; {\rm d}x'\;  \phi_-(x', s_n) S(x',s_n) =\nonumber \\
&=\frac{c_{0{\rm L}}^2}{(s_n-2s_0) W'(s_n)} \int_{-\infty}^{+\infty} \; {\rm d}x'\;  \phi_-(x', s_n) \; h(x'),
\end{align}
and we can pick again only the dominant oscillatory contribution to write
\begin{equation}
    \chi_2^S(x,t)\Big|_{W \text{ poles}} = c_W \;e^{ts_0}\; \phi_-(x,s_0).
\end{equation}
We stress again that in general the $s_0$ found at second-order could be different from the $s_0$ found at first-order, because it could be characterized by an angular momentum $l$ different from the one at first-order. This is a crucial point and we will come back to it in the next section. 
More importantly, the solution at second-order has  a contribution which oscillates as $2s_0$, being $s_0$ 
the linear frequency. 

\begin{figure}[h!]
    \centering
    \includegraphics[scale=0.6]{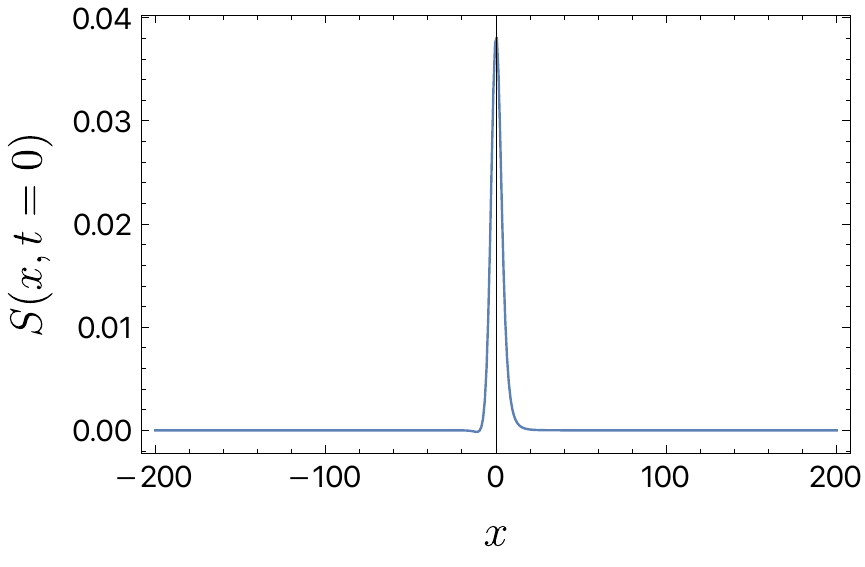}
    \caption{Numerical evaluation of the initial source in the specific case of $l=2$ and $m=0$, with head-on initial conditions, as in Ref. \cite{Nicasio:1998aj}. For the plot we used units of $M=1$.}
    \label{fig:source}
\end{figure}

The last step is to  take now the limit $x\to \infty$ for the source term at $2s_0$. Here comes our basic assumption. If the source is localised in a region $x'_L< x'<x'_R$, we can assume that we are  always in the case $x>x'$. This assumption is justified because we know that, for the QNMs,  the light ring region dominates in the source at the initial time (aftwerwards, the source propagates),  see in Fig. \ref{fig:source}, and the  observers are far away from it. Hence we can write in this limit (an alternative derivation is offered in  \ref{sec_another_derivation})
\begin{align}
\label{int}
\chi_2^S(x,t)\Big|_{S \text{ pole}} = c_{0{\rm L}}^2 e^{2s_0 t} \; \phi_+(x,2s_0) \; \int_{-\infty}^{+\infty} \; {\rm d}x'\;  \frac{\phi_-(x', 2s_0)}{W(2s_0)}\; h(x'),
\end{align}
which, defining
\begin{equation}
\label{cs}
    c_{S}=  \frac{c_{0{\rm L}}^2}{W(2s_0)} \; \int_{-\infty}^{+\infty} \; {\rm d}x'\;  \phi_-(x', 2s_0)\; h(x'),
\end{equation}
can be rewritten as
\begin{equation}
\label{a}
    \chi_2^S(x,t)\Big|_{S \text{ pole}}= c_S \;e^{2s_0 t} \; \phi_+(x,2s_0).
\end{equation}
Recalling the shape of $\phi_-$,  it might seem that this integral is divergent as it grows exponentially at infinity,   a typical behaviour when studying QNMs only on the $x$ variable, as also pointed out in \cite{08article}. The convergent behaviour is ensured by causality, i.e. reintroducing the time variable $t$ and a theta function $\theta(t-x)$. We have not  introduced it  for the sake of clarity of our equations, and because in our estimate we will take only the peak of the source, assuming its convergence.
We get a form which resembles the linear case one. It is this property  which will allow us to considerably simplify the non-linearities of the QNMs.

\FloatBarrier

\section{The dependence on the initial conditions}
\label{sec_initcond}
We are now in the position to offer some considerations about the dependence on the initial conditions. As we have seen, the second-order solution has various contributions. 

\begin{enumerate}

\item The first one $\chi_2^I$ comes from the solution of the homogeneous equation. Depending upon the angular momentum we are interested in, this can  renormalize  or not  the first-order solution, i.e. in the specific non-linearity ratio, the linear frequency part at the denominator will receive contributions also from the non-linear part.\\
For instance, one can consider an $l=4$  source term  generated from the linear solution at  $l=m=2$. This was the case   considered in Refs. \cite{Mitman:2022qdl, Cheung:2022rbm}. In such a case, the Zerilli potentials are different at first- and second-order and therefore the solutions of the homogeneous equations
are different and the linear solution is not renormalized.

\item The second contribution $\chi_2^S\Big|_{W \text{ poles}}$ comes from the source term picking up only the poles of the Wronskian and again the corresponding frequency can be equal or different from the linear one, thus it will renormalize or not the linear solution.

\item Finally, the third contribution $\chi_2^S\Big|_{S \text{ pole}}$ comes from the source term picking up a frequency which is twice the linear frequency.

\end{enumerate}
Let us consider the case in which the source is constructed from an angular momentum which is different from the linear solution, like in Refs. \cite{Mitman:2022qdl, Cheung:2022rbm}.
If so, extracting from the second-order solution only the third contribution and comparing to the first-order solution at null infinity, we get

\begin{equation}
	{\rm NL}=\left|\frac{ \psi_2(x\rightarrow \infty,2s_0)}{\psi^2_1(x\rightarrow \infty,s_0)}\right|=\left|\frac{1}{2s_0}\frac{c_S}{c_{0{\rm L}}^2}\frac{\phi_+(x\rightarrow \infty,2s_0)}{\phi^2_-(x\rightarrow\infty,s_0)}\right|,
\end{equation}
where we used the asymptotic relation 
\begin{equation}
    \psi_2(x\to \infty, t) = \frac{1}{2s_0}\chi_2(x\to \infty, t),
\end{equation}
which shows that the ratio does not depend on the initial condition of the problem as $c_S$ scales like $c_{0{\rm L}}^2$, see Eq. (\ref{cs}). This explains the mild dependence on the initial conditions found in Ref. \cite{Cardoso_new}. Notice also that at null infinity $\phi_+(x,2 s_0)\sim \phi^2_-(x,s_0)\sim e^{-2s_0 x}$ and the dependence on $x$ correctly cancels out.

Let us summarize here the basic results which might a bit obscured by the many expressions we have written so far:

\begin{enumerate}
    
    \item If the second-order source is well localized and the  homogeneous solution of the second-order mode has a frequency which is different from  the frequency of the linear mode, then defining NL by extracting from the second-order result the piece with such double frequency, the ratio NL depends mildly on the initial conditions of the problem at null infinity. Furthermore, the amplitude of the second-order contribution scales indeed as the square of the amplitude of the linear one.

    \item Still under the assumption of a well localized source, if the  homogeneous solution of the  second-order mode has a frequency which is equal to  the frequency of the linear mode, then defining NL by extracting from the second-order result the piece with such double frequency, the ratio NL depends on the initial conditions of the problem. Such a dependence becomes mild when the renormalization of the linear piece from the second-order homogeneous solution is negligible.

\item The amount of non-linearity will depend on some intrinsic properties of the final black hole state, such the mass and the spin. They will show up in the second-order source and the non-linearity will automatically depend on it.
    
\end{enumerate}
\section{Evaluating the non-linearity from the WKB approximation}
\label{sec_WBKnonlin}
In order to estimate the level of second-order non-linearities, we can use the WKB approximation which we summarize in \ref{sec_wkb}. Our work differs from Ref. \cite{Lagos:2022otp} mostly because we are not using a toy model source, but a realistic one to give the estimate. Furthermore we are interested in explicitly evaluate the nonlinearity predicted in Ref. \cite{Cheung:2022rbm}.
We start from the crucial
 term (\ref{cs}), which we rewrite as
 
\begin{equation}
    I_l\equiv\int_{-\infty}^{+\infty} \; {\rm d}x'\; \phi_{-}^{(l)}(x', 2s_0 )\; h^{(l)}(x'),
\end{equation}
where we have introduced the angular momentum $l$ to keep track of the different solutions and sources
\begin{equation}
h^{(l)}(x)=\sum_{i=0}^{N_{\rm max}} R_i(r(x))\; (\partial_x^{m_i} \phi_-^{(l_1)} (x,s_0) )\; (\partial_x^{n_i} \phi_-^{(l_2)} (x,s_0)).
\end{equation}
From the WKB analysis of the linear solution we can provide an estimate for $h_l(x)$. 
The  source is peaked  close the potential peak, and is therefore a well localized function in the region $I_2$ of the WKB, defined in \ref{sec_wkb}, where $\phi_-$ can be written as an exponential, as in Eq. (\ref{midphi-}). 
 
 We focus  on the case 
 \begin{equation}
     n=0, \quad l_1=l_2=\frac{l}{2}
 \end{equation}
for which the linear solution is not renormalized. The procedure can be also followed in the case in which the linear solution is renormalized, and the result have a similar form.
We get (assuming a mild dependence of $x_0$ on the multipoles)
\begin{equation}
    \phi_{-}^{(l_1)} = A_-(l_1) \;e^{-i \sqrt{k_{l_1}}(x-x_0)^2/2}, \quad x\in I_2,
\end{equation}
where $k_{l_1}$ is defined in Eq.  (\ref{ktdef}) and $A_-(l)$ is evaluated for the specific $l$ and for $s_{(l,\,0)}$, which in this case is $s_0$, found at linear level. The region $I_2$ in the WKB approximation is identified as the one between the two inversion points, close to the potential peak. Those regions $I_{1,2,3}$ are fully defined in the \ref{sec_wkb}.

We can estimate the integral using the method of the steepest descent close to the point $x=x_0$.
Being $\phi_{-}^{(l_1)}$ an exponential, we can factor out the function and write $h^{(l)}(x)$ as
\begin{equation}
    h^{(l)}(x) = A_-(l_1)^2 \;e^{-i \sqrt{k_{l_1}}(x-x_0)^2} H^{(l)}(x),
\end{equation}
where we defined for simplicity the rational function
\begin{equation}
    H^{(l)}(x)=\sum_{i=0}^{N_{\rm max}} R_i(r(x))\; P_{m_i}(x)\; P_{n_i}(x),
\end{equation}
and $P_{n_i}(x)$ and $P_{m_i}(x)$ are polynomials of degree $n_i$ and $m_i$.
Substituting in the integral we get
\begin{align}
   I_l=A_-(l) \; A^2_-(l_1)  \int_{-\infty}^{+\infty} \; {\rm d}x'\;  e^{-i \sqrt{k_{l}}(x'-x_0)^2/2} \; \;e^{-i \sqrt{k_{l_1}}(x'-x_0)^2} \;H^{(l)}(x'),
\end{align}
and the steepest descent method can be immediately evaluated by taking $x'=x_0$ inside the polynomial terms and computing the resulting phase
\begin{align}
    I_l=A_-(l)  A^2_-(l_1)   e^{-i\pi/4}\;\sqrt{\frac{2\pi}{\sqrt{k_l}+ 2\sqrt{k_{l_1}}}}\;
   H^{(l)}(x_0).
\end{align}
The non-linearity parameter $\rm NL$  is then extracted from

\begin{align}
    \frac{\psi_2^{(l)}}{\left(\psi_1^{(l_1)}\right)^2}=&\Bigg|\frac{1}{2s_0} \frac{\phi_+(x\rightarrow \infty,2s_0)}{\phi^2_-(x\rightarrow\infty,s_0)}\frac{A_-(l)  A^2_-(l_1)   e^{-i\pi/4}}{W^{(l)}(2s_0)}\;\times \nonumber \\
    &\times \;\sqrt{\frac{2\pi}{\sqrt{k_l}+ 2\sqrt{k_{l_1}}}}\;H^{(l)}(x_0) \Bigg|.
\end{align}
Using the approximate form for the Wronskian in WKB and the solution $\phi_-$ extended to $+\infty$, defined in  \ref{sec_wkb} we get

\begin{align}
\label{NLWKB}
    \frac{\psi_2^{(l)}}{(\psi_1^{(l_1)})^2} &= \Bigg|\frac{1}{2\sqrt{2s_0}}\,\frac{ 2^{-\bar{\nu}/2}}{(4k_{l})^{1/4}}  \; \Gamma\left(\frac{-\bar{\nu}}{2}\right) \;\frac{A_+(l_1)^2}{ A_+(l)}\;\times \nonumber \\
    &\times \;\sqrt{\frac{2}{\sqrt{k_l}+ 2\sqrt{k_{l_1}}}}\;H^{(l)}(x_0) \Bigg|
\end{align}
where $\bar{\nu}$ satisfies \cite{Schutz:1985km}
\begin{equation}
     \bar{\nu} = i\; \frac{(2s_0)^2 + V_l(x_0)}{\sqrt{2Q_l''(x_0)}} -\frac 12, \quad \bar{\nu} \notin \mathbb{N}.
\end{equation}
We can notice that most of the coefficients connecting the various pieces of the solution simplify in the end and we are left with the ratio between the matching of the regions $I_2$ and $I_3$, $A_+(l_1)$ at linear level and the same matching coefficient but at nonlinear level $A_+(l)$. They are all defined in \ref{sec_wkb}.

\section{Examples}
We can provide now a few example. We will start with a source term for the second-order with $l=4$, generated from a linear solution with $l_1=l_2=2$. We are therefore in the case in which  no renormalization of the linear coefficient takes place and the non-linearity does not depend on the initial conditions. In this case we can immediately apply the formulas derived above, in particular  Eq. (\ref{NLWKB}), after having evaluated the source term at the peak of the potential. 

We use the source obtained in Ref. \cite{Nakano:2007cj}, reported  in \ref{sec_sources}, Eq. (\ref{Snakano}), valid for a Schwarzschild BH. We  notice that the only dependence on $m$ that we obtain is in the source term, which is generated with a linear term $l_1=l_2=2$ and $m_1=m_2=2$.
We  get
\begin{equation}
	{\rm NL}=\left|\frac{\psi_2^{l=4,\,m=4}}{\left(\psi_1^{l_1=2,\,m_1=2}\right)^2}\right| \simeq 0.51.
\end{equation}
If we instead calculate the non-linearity of the GW strain $h$, accounting for the corresponding 
normalization $\psi=2\ddot h$ and accounting for the two polarization, we get, in units of the BH mass $M$

\begin{equation}
{\rm NL}_h= \omega_{220}^2{\rm NL}\simeq 0.069,\,\,\,\omega_{220}\simeq 0.37,
\end{equation}
which fits well the result in Ref. \cite{Cardoso_new} (see their Fig. 3).

In the case in which the  non-linearity is evaluated for $l=2$ and $l_1=2$, we follow a similar procedure with respect to the one presented more generally in section \ref{sec_WBKnonlin}. The main differences are the solutions $\phi_{\pm}$, which now are the same as the one found at linear level. So $A_{\pm}$ are the same for the two orders and $k_{l_1}=k_{l_2}=k_l$.
For such a case, we  consider the source reported in \ref{sec_sources}, Eq. (\ref{Sgleiser}), derived  in Ref. \cite{Nicasio:1998aj} for the head-on collision of two  Schwarzschild BHs resulting in a spinless BH. The two BHs start very close to each other at a distance $L$ and with momentum $P$, as described in the series of works \cite{Price:1994pm, Gleiser:1995gx, Gleiser:1996yc, Nicasio:1998aj}.
This behaviour is encoded in the initial conditions, which is Misner's wormhole solution \cite{Misner:1960zz}, where the two BH solution is constructed as a wormhole solution at fixed time. The corresponding starting conditions for the linear solution are
\begin{align*}
\tag{\stepcounter{equation}\theequation}
    &\psi_1(r,\,t=0) =  \frac{8}{3} \frac{ML^2 (5 \sqrt{r-2M} + 7 \sqrt{r})r}{(\sqrt{r}+ \sqrt{r-2M})^5 (2r+3M)},\\
    &\dot{\psi}_1(r,\,t=0)  = \frac{\sqrt{r-2M}\, P L (8r+6M)}{r^{5/2}(2r+3M)},
\end{align*}
and have a more complicated expression for the non-linear solution \cite{Nicasio:1998aj}.
Such initial conditions are the evolved using the usual  BH theory through the  Zerilli potential.
The form of the non-linearity in this case can be written as
\begin{align*}
{\rm NL}&=\left|\frac{ \psi_2(x\rightarrow \infty,t)}{\psi^2_1(x\rightarrow \infty,t)}\right|= \\
&=\frac{\left| c^2_{0{\rm L}} \,\int_{-\infty}^{+\infty} \; {\rm d}x'\;  \frac{\phi_-(x', 2s_{2,0})}{W(2s_{2,0})}\; h_{20\times 00}(x')\right|}{\left| c_{0{\rm L}} + c_W - \frac{c^2_{0{\rm L}}}{s_{2,0} W'(s_{2,0})} \int_{-\infty}^{+\infty} \; {\rm d}x'\;  \phi_-(x', s_{2,0}) \; h_{20\times 00}(x') \right|^2} \times
\\
&\times  \left| \frac{1}{2s_0} \frac{\phi_+(x\rightarrow \infty,2s_0)}{\phi^2_-(x\rightarrow\infty,s_0)}\right|.
\tag{\stepcounter{equation}\theequation}
\end{align*}
Inserting the various parameters in the formula above we obtain a non-linearity equal to
\begin{equation}
    {\rm NL} = \left| \frac{\psi_2^{l=2,\,m=0}}{\left(\psi_1^{l_1=2,\,m_1=0}\right)^2}\right| \simeq 0.06,
\end{equation}
which is much smaller than the one generated at $l=4$ found above. We have numerically checked that the dependence on the initial conditions is rather mild, being the renormalization of the linear solution basically absent.

\section{Conclusions}
Non-linearities in QNMs are a relevant subject as it will help to improve our understanding of the BH ringdown and probe the non-linear nature of gravity.  In this paper we have described a perturbative second-order method to estimate analytically the non-linearity for the QNMs generated at the ringdown of a BH. The method relies only on the knowledge of the specific source and on the fact that it is well peaked, typically around the light ring, where the potential has a maximum. The amplitude of the second-order contribution scales like the square of the amplitude of the linear contribution and we have described under which conditions the non-linearity depends only mildly on the initial conditions. Our perturbative approach has certainly limitations due to its inability to account for the  back-reaction of the GWs onto the BH geometry. 

Furthermore, 
we have analyzed  the propagation of  the  linear and quadratic BH perturbations in the eikonal limit and, by using the WKB approximation, we have obtained  solutions to the linear and quadratic perturbation equations. However,
the implemented WKB approximation works better in the large multipole limit, while the GW signals are dominated by low-frequency waves. It could be possible to extend this work further by trying to apply the same techniques to a spinning BH, which could complement the analysis and allow us to compare also with other results present in the literature.
\vskip 0.5cm
\noindent
\section*{Acknowledgments}
 We thank E. Berti, B. Bucciotti, V. Cardoso, G. Carullo, R.J. Gleiser, C.O. Nicasio, A. Pound, R.H. Price, J. Pullin, J. Redondo–Yuste, J.L. Ripley, A. Spiers, and B. Wardell for many useful discussions. The work of D. P.   is supported by the Swiss National Science Foundation under grants no. 200021-205016 and PP00P2-206149. A.K. is supported by the PEVE-2020 NTUA programme for basic research with project number 65228100. A. K. thanks the University of Geneva where part of this work was performed. A.R. is funded by the Boninchi Foundation.

\newpage

\appendix

\section{Sources}
\label{sec_sources}
Here we list all the sources used to compute the numerical estimates of non-linearities, already assuming that the solution at linear order behaves like $e^{s_0 t}$, so that we can factor out the temporal behaviour.
The first one is $l=2, \; m=0$ source, taken from Ref. \cite{Nicasio:1998aj} for a Schwarzschild BH, where the $\psi$ is the solution at linear order for $l=2$ and $m=0$, and the subscripts represents derivatives with respect to the $r$ variable.

The second one is $l=4, \; m=4$ source, taken from Ref. \cite{Nakano:2007cj} for a Schwarzschild BH, where the source is built from the linear solution $\psi$ with $l=2$ and $m=2$

Both sources have been regularised in the two cited works and both go to zero at infinity and at the horizon of the BH
\\
\begin{strip}
\begin{align*}
\label{Sgleiser}
    S_{l=2,\,m=0}(r,t)=e^{2s_0 t} \;\frac{12 (r-2)^3}{7 (2 r+3)} \Bigg( \frac{(r-2) (2 r+3) \psi_{rr}^2}{3 r^4}-\frac{s_0 \psi  \psi_{rr}}{r^3 (2 r+3)}+\frac{(3 r-7) s_0 \psi _r^2}{3 (r-2) r^3}+\frac{s_0 \psi _r \psi_{rr}}{3 r^2}\\-\frac{s_0 \psi  \psi_{rrr}}{3 r^2}
-\frac{(2 r+3) s_0^2 \psi _r^2}{3 (r-2) \
r^2}+\frac{4 \left(3 r^2+5 r+6\right) \psi _r \psi_{rr}}{3 r^5}-\frac{12 \left(r^2+r+1\right)^2 \
s_0^2 \psi ^2}{(r-2)^2 r^4 (2 r+3)}+\\\frac{\left(8 r^2+12 r+7\right) s_0 \psi  \psi _r}{(r-2) r^4 (2 \
r+3)}-\frac{\left(2 r^2-1\right) s_0 \psi  \psi _{rr}}{(r-2) r^3 (2 r+3)}-\frac{4 \
\left(r^2+r+1\right) s_0^2 \psi  \psi _r}{(r-2)^2 r^3}-\\\frac{4 \left(2 r^3+4 r^2+9 r+6\right) \psi  \psi_{rr}}{r^6 (2 r+3)}+
\frac{\left(12 r^3+36 r^2+59 r+90\right) \psi _r^2}{3 (r-2) \
r^6}+\frac{\left(18 r^3-4 r^2-33 r-48\right) s_0 \psi  \psi _r}{3 (r-2)^2 r^4 (2 r+3)}+\\ \frac{\left(112 \
r^5+480 r^4+692 r^3+762 r^2+441 r+144\right) s_0 \psi ^2}{(r-2)^2 r^5 (2 r+3)^3}+
\\ \frac{12 \left(2 r^5+9 \
r^4+6 r^3-2 r^2-15 r-15\right) \psi ^2}{(r-2)^2 r^8 (2 r+3)}-\frac{2 \left(32 r^5+88 r^4+296 r^3+510 \
r^2+561 r+270\right) \psi  \psi _r}{(r-2) r^7 (2 r+3)^2}\Bigg).
\tag{A1}
\end{align*}
\end{strip}
\\

\clearpage
\begin{strip}
\begin{align*}
\label{Snakano}
S_{l=4,\,m=4}(r,t)=-e^{2s_0 t}\;\frac{1}{9} \sqrt{\frac{5}{14 \pi }} s_0 \Bigg(\Bigg(\frac{(r-2) (7 r+4) \
s_0^2}{r}-\frac{3 (r-2)}{r^5 (2 r+3)^2 (3 r+1)^2} (228 r^7+ \\
+8 r^6-370 r^5+142 r^4-384 r^3-514 r^2-273 r-48)
\Bigg) \psi_r^2\\
+\psi \psi_r  \Bigg(\frac{4 (r-2)^2 s_0^2}{r^2 (3 r+1)^2}+\frac{6 (r-2)}{r^6 (2 r+3)^3 (3 r+1)^2} (144 r^8+4116 r^7+2154 r^6-2759 r^5-8230 r^4-\\
9512 r^3-3540 r^2-1119 r-144)\Bigg) \\
+\psi^2 \Bigg(-\frac{3 (276 r^7+476 r^6-1470 r^5-1389 r^4-816 r^3-800 r^2-555 r-96) s_0^2}{(r-2) r^3 (2 r+3)^2 (3 r-1)^2}+
\\
\frac{9 (r-2)}{r^7 (2 r+3)^4 (3 r+1)^2} (2160 r^9+11760 r^8+30560 r^7+\\
41124 r^6+31596 r^5+11630 r^4-1296 r^3-4182 r^2-1341 r-144)-\frac{r (7 r+4) s_0^4}{r-2}\Bigg)
\Bigg).
\tag{A2}
\end{align*}
\end{strip}

\section{Comments on the non-linear overtones in the source}
In this work we have used heavily the approximation that at linear level only the first overtone $n=0$ matters,
\begin{equation}
    \psi = \sum_n c_n \; \phi(x,s_n)\; e^{ts_n} \sim c_{0{\rm L}} \; \phi(x,s_0)\; e^{ts_0}.
\end{equation}
However in the source we have a quadratic dependence on the first-order solution, so the sum squared will contain several overtones. We show here that  they do not count  in the case of a non-linearity evaluated as in Ref. \cite{Mitman:2022qdl}. This is  because of the pole structure of the Laplace transformed source.
Indeed, we have seen that in general the source will have the form
\begin{equation}
S(x,t) = \sum_{i=0}^{N_{\rm max}} R_i(r(x)) \left(D_{i,1}[t, x] \psi_1 (x,t) \right) \left(D_{i,2}[t, x] \psi_1 (x,t) \right).
\end{equation}
If we plug in the full linear solution $\psi_1$ we get
\begin{align}
    S(x,t) = \sum_{n} \sum_{m} c_n \, c_m \, e^{t (s_n + s_m)} \;\times \nonumber\\
    \times \;\sum_{i=0}^{N_{\rm max}} R_i(r(x)) \left(\partial_x^{k_i}\; \phi (x,s_n) \right) \left(\partial_x^{j_i} \; \phi (x,s_m) \right).
\end{align}
For the sake of clarity, we relabel 
\begin{equation}
    h_{nm}(x) = \sum_{i=0}^{N_{\rm max}} R_i(r(x)) \left( \partial_x^{k_i}\; \phi (x,s_n) \right) \left(\partial_x^{j_i} \; \phi (x,s_m) \right)
\end{equation}
and the source term becomes
\begin{equation}
    S(x,t) = \sum_{n} \sum_{m} c_n \, c_m \, e^{t (s_n + s_m)} h_{nm}(x).
\end{equation}
As we Laplace transform we get simply
\begin{equation}
    S(x,s) = \sum_{n} \sum_{m} \frac{ c_n \, c_m \,}{s-(s_n + s_m)} h_{nm}(x)
\end{equation}
so that   the pole structure will enter in the "renormalisation" of the linear order, but only the selected pole $1/(s-2s_0)$ will contribute to the dominant frequency non-linearity asymptotically.

\section{Another derivation of Eq. (\ref{a})}
\label{sec_another_derivation}
Consider the case in which the first-order solution is constructed from an $l=2$ mode and the second-order from an $l=4$ mode. The solution at second-order $\chi_2^{l=4}(x,t)$
satisfies 

\begin{equation}
( \partial^2_x - \partial^2_t - V(x) )\,\chi_2^{l=4} (x,t) = S_0(x) e^{t\, 2s_{2,0}}, 
\label{ss2}
\end{equation}
where, we have use Eq. (\ref{ss1}) with 
$S_0(t)= c^2_{0{\rm L}}\; 
h(x)$.
It is then easy to verify that the general solution to Eq. (\ref{ss2}) is 
\\

\begin{align}
\label{ss0}
\chi_2^{l=4} &\simeq A_{4,+}^{(2)} \,\phi_{+,l=4}\,(x,s_{4,0}) e^{t s_{4,0}} + A_{4,-}^{(2)} \,\phi_{-,l=4}\,(x,s_{4,0}) e^{t s_{4,0}} + \nonumber \\
&+e^{t\,2s_{2,0} }
\Psi_2^{l=4}(x,2s_{2,0}),
\end{align}
where $\phi_{\pm,l=4}\,(x,s_{4,0}) $ are the two in independent  solutions of the equation 
\begin{equation}
( \partial^2_x - s_{4,0}^2 - V(x) )\,\phi_{\pm,l=4}\,(x,s_{4,0}) = 0,
\label{ss3}
\end{equation}
and $\Psi_2^{l=4}(x,2s_{2,0})$ is a particular integral of the equation 
\begin{equation}
( \partial^2_x - 4s_{2,0}^2 - V(x) )\,\Psi_2^{l=4}(x,2s_{2,0}) = S_0(x).
\label{ss4}
\end{equation}
Therefore, the only unknown at second-order is the particular integral $\Psi_2^{l=4}(x,2s_{2,0})$ of Eq. (\ref{ss4}). To determine the latter, we use the method of the variation of parameters for solving non-homogeneous differential equations where the particular integral is expressed as a linear combination of the homogeneous problem, but with $x$-depended  coefficients. We write 
\begin{equation}
\label{ss5}
    \Psi_2^{l=4}(x,2s_{2,0})=u_+(x) \phi_{+,l=4}(x,2s_{2,0})+
    u_-(x) \phi_{-,l=4}(x,2s_{2,0}),
\end{equation}
where $\phi_{\pm,l=4}(x,2s_{2,0})$ solves the homogeneous equation
\begin{equation}
\label{ss6}
( \partial^2_x - 4s_{2,0}^2 - V(x) )\,\phi_{\pm,l=4}(x,2s_{2,0}) = 0.
\end{equation}
Then imposing the condition 
\begin{eqnarray}
\label{ss7}
    u_+' \phi_{+,l=4}+
    u_-' \phi_{-,l=4}=0,
\end{eqnarray}
we find from Eq. (\ref{ss4}) 
\begin{eqnarray}
\label{ss8}
     u_+' \phi_{+,l=4}'+
    u_-' \phi_{-,l=4}'=S_0(x).
\end{eqnarray}
Solving Eqs.(\ref{ss7}) and (\ref{ss8}) we find that
\begin{eqnarray}
    u_+'&=&\frac{S_0(x) \phi_{-,l=4}(x,2s_{2,0})}
    {\phi_{+,l=4}\phi_{-,l=4}'-\phi_{-,l=4}\phi_{+,l=4}'}= \nonumber\\
    &=&\frac{S_0(x) \phi_{-,l=4}(x,2s_{2,0})}
    {W(2s_{2,0})},\nonumber \\
     u_-'&=&-\frac{S_0(x) \phi_{+,l=4}(x,2s_{2,0})}
    {\phi_{+,l=4}\phi_{-,l=4}'-\phi_{-,l=4}\phi_{+,l=4}'}=\nonumber\\
    &=&-\frac{S_0(x) \phi_{+,l=4}(x,2s_{2,0})}
    {W(2s_{2,0})}.
\end{eqnarray}
A simple integration  then gives 
\begin{eqnarray}
    u_+&=&\int_{-\infty}^x \; {\rm d}x'\;
    \frac{ c^2_{0{\rm L}}   \; 
h(x')\phi_{-,l=4}(x',2s_{2,0})}
    {W(2s_{2,0})}\approx\\
    &\approx& \int_{-\infty}^\infty \;  {\rm d}x'\;
    \frac{ c^2_{0{\rm L}}   \; 
h(x')\phi_{-,l=4}(x',2s_{2,0})}
    {W(2s_{2,0})},
    \nonumber \\
     u_-&=&-\int_x^{\infty}\; {\rm d}x'\;\frac{ c^2_{0{\rm L}}  \; 
h(x')\phi_{+,l=4}(x',2s_{2,0})}
    {W(2s_{2,0})}\approx 0,
\end{eqnarray}
where we have used the fact that $h(x)$ is a localized source and we're interested in values of $x\to \infty$. Therefore, we get that the partial integral is given by

\begin{eqnarray}
     &\Psi_2^{l=4}(x,2s_{2,0})\approx\left\{  \int_{-\infty}^\infty \; {\rm d}x' \;
    \frac{ c^2_{0{\rm L}}  \; 
h(x')\phi_{-,l=4}(x',2s_{2,0})}
    {W(2s_{2,0})}\right\} \times \nonumber \\
    &\times \;\phi_{+,l=4}(x,2s_{2,0})\nonumber \\
    &= \bar{c}_S \phi_{+,l=4}(x,2s_{2,0}),
\end{eqnarray}
which reproduces  Eq.  (\ref{a}).

\section{The WKB approximation}
\label{sec_wkb}
 In this Appendix we follow Ref. \cite{Schutz:1985km} using WKB techniques to find the Green's function.
We start again from the equation
\begin{equation}
   \left(\partial_x^2  - s^2 - V(x) \right)\; \phi_{\pm}=0.
\end{equation}
The QNMs can be understood as solutions where the transmission and reflection of the wave have approximately the same amplitude. So we expect to be able to find those solutions by considering "energies" of order of the maximum of the potential.\\
This means that we require, around the maximum $x_0$ of $V(x)$
\begin{equation}
    s^2 + V(x) \simeq 0,
\end{equation}
and, expanding $V(x)$
\begin{equation}
    s^2 + V(x_0) + \frac 12 V''(x_0)(x-x_0)^2,
\end{equation}
we need that
\begin{equation}
    s^2 + V(x_0) > 0, \quad V''(x_0) < 0.
\end{equation}
This expansion in a quantum mechanical  point of view means that we are considering ``energies" just below the peak of the potential, so we can identify two inversion points, $x_1$, $x_2$, located at
\begin{equation}
    s^2 + V(x)=0,
\end{equation}
which will result in two inversion points, located at $x_1$ and $x_2$.
We can then identify three regions  $I_1 = (-\infty, x_1)$, $I_2 = (x_1, x_2)$,
     $I_3 = (x_2,\infty)$, 
as shown in Fig. 1.
Let us define 
\begin{equation}
    Q(x,s) = - s^2 - V(x),
\end{equation}
to rewrite the problem in a WKB fashion 
\begin{equation}
   \partial_x^2   \phi = -Q(x,s) \phi.
\end{equation}

\begin{figure}
    \centering
    \includegraphics[scale=0.6]{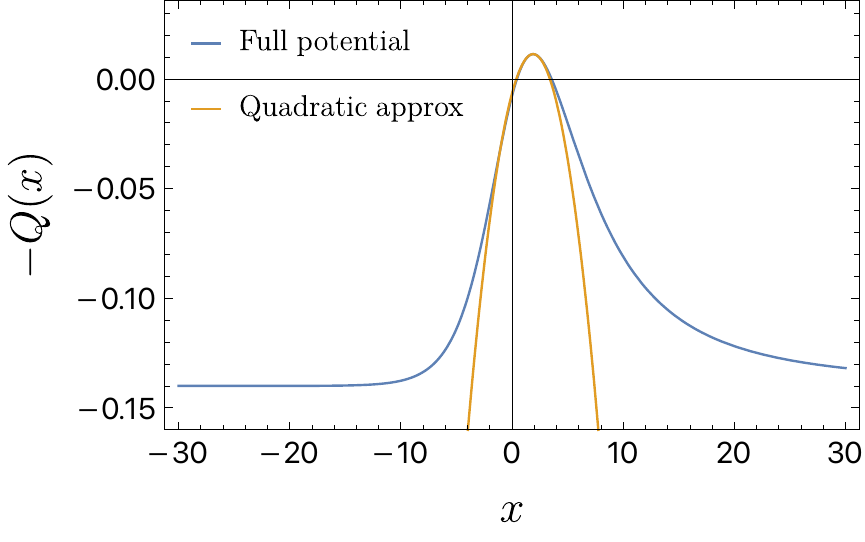}
    \caption{$Q(x)$ and its quadratic approximation, The various regions are identified by the solutions of $Q(x)=0$, as done in Ref. \cite{Schutz:1985km}. The plot has units $M=1$.}
    \label{fig:pot}
\end{figure}

In each of the regions certain approximations hold, so we have
\begin{itemize}
    \item region 1: $I_1 = (-\infty, x_1)$, here the solution is standard WKB up to the inversion point, namely
    \begin{equation}
        \phi^{(1)}(s,x) = \frac{1}{Q(x,s)^{1/4}}\exp{\left(i\int_{x}^{x_1} \sqrt{Q(x',s)}\;{\rm d}x'\right)},
    \end{equation}
    which asymptotically behaves like\footnote{We have to pick the first branch of $s$ $\lim_{\epsilon\to 0} s + i\epsilon$ to be consistent with the sign choices.}
    \begin{equation}
        \lim_{x\to-\infty}\phi^{(1)}(x,s) \sim \frac{1}{s^{1/2}}e^{-s(x_1-x)}\sim e^{sx}.
    \end{equation}

     \item region 3: $I_3 = (x_2,\infty)$, here we have almost the same solution with respect to region~1
    \begin{equation}
        \phi^{(3)}(s,x) = \frac{1}{Q(x,s)^{1/4}}\exp{\left(i\int_{x_2}^{x} \sqrt{Q(x',s)}\;{\rm d}x'\right)},
    \end{equation}
    which asymptotically behaves like
    \begin{equation}
        \lim_{x\to \infty}\phi^{(3)}(x,s) \sim \frac{1}{s^{1/2}}e^{-s(x-x_2)}\sim e^{-sx}.
    \end{equation}

       \item region 2: $I_2 = [x_1, x_2]$, here we have to solve the full differential equation, but we can approximate the potential by expanding it up to second-order
    \begin{equation}
        \left(\partial_x^2 - s^2 - V(x_0) - \frac 12 V''(x_0)(x-x_0)^2\right)\; \phi^{(2)} =0.
    \end{equation}
    We can make the following substitutions to put the equation in a known form
    \begin{equation}
    \label{ktdef}
        k=-\frac 12 V''(x_0) = \frac 12 Q''(x_0), \quad t=(4k)^{1/4} e^{i\pi/4} (x-x_0), 
    \end{equation}
    \begin{equation*}
        \nu + \frac{1}{2} = -i\frac{Q(x_0)}{\sqrt{2Q''(x_0)}} = i \frac{(s^2 + V(x_0))}{\sqrt{-2V''(x_0)}} 
    \end{equation*}
    and the equation becomes
        \begin{equation}
        \left(\partial_t^2 + \nu + \frac{1}{2} - \frac{1}{4}t^2\right)\; \phi^{(2)}(t) =0,
    \end{equation}
    which has known solutions in terms of parabolic cylinder functions
    \begin{equation}
        \phi^{(2)}(t) = A D_{\nu}(t) + B D_{-1-\nu}(it).
    \end{equation}
The parabolic cylinder function $D_{k}(t)$, with $k \in \mathbb{N}$, can be also rewritten as
\begin{equation}
    D_{k}(t) = 2^{-k/2} e^{-t^2/4} H_{k} \left(\frac{t}{\sqrt{2}} \right),
\end{equation}
where $H_{\nu}$ is an Hermite polynomial.
    The QNM solution from matching implies a discrete set of frequencies, given by
    \begin{equation}
    \label{qnm_easy}
        \frac{Q(x_0)}{\sqrt{2Q''(x_0)}} = i\left(n + \frac{1}{2}\right),\quad n \in \mathbb{N},
    \end{equation}
    with the two additional conditions
    \begin{equation}
        B=0, \quad \nu = n.
    \end{equation}
\end{itemize}

\subsection{Transmission and matching}
Let us be more precise about the functional forms of $\phi_{\pm}$ and review the matching procedure.
The main property that we will use is that the expansion of the parabolic cylinder function solution behaves at $\pm \infty$ differently 
\begin{align}
\label{pluslim}
    \lim_{x\to +\infty}\phi^{(2)}(x) =   B e^{-3i\pi (\nu+1)/4} (4k)^{-(\nu+1)/4} (x-x_0)^{-(\nu+1)}e^{i \sqrt{k}(x-x_0)^2} +\\
    \left[ A + B \frac{\sqrt{2\pi} e^{i \nu \pi /2}}{\Gamma(\nu + 1)}\right]e^{i \pi \nu /4} (4k)^{\nu/4} (x-x_0)^{\nu}e^{-i \sqrt{k}(x-x_0)^2},
\end{align}
\begin{align}
\label{minuslim}
    \lim_{x\to -\infty}\phi^{(2)}(x) =   A e^{-3i\pi \nu/4} (4k)^{\nu/4} (x-x_0)^{\nu}e^{-i \sqrt{k}(x-x_0)^2} +\\
    \left[ B -i A \frac{\sqrt{2\pi} e^{-i \nu \pi /2}}{\Gamma(-\nu )}\right]e^{i \pi (\nu+1) /4} (4k)^{-(\nu+1)/4} (x-x_0)^{-(\nu+1)}e^{i \sqrt{k}(x-x_0)^2},
\end{align}
and we are interested only in the term $\exp(-i\sqrt{k}(x-x_0)^2)$, cancelling the other oscillatory behaviour.
Let us start from the matching procedure to find $\phi_{\pm}$ in the central region and then we'll match to find the whole solution. At $x\to -\infty$ we are in the $I_1$ region, so we have
\begin{equation}
    \phi^{(1)}(s,x) = \frac{1}{Q^{1/4}(x,s)}\exp{\left(i\int_{x}^{x_1} \sqrt{Q(x',s)}\;{\rm d}x'\right)}
\end{equation}
and to match it with $\phi^{(2)}$ we need to impose both the conditions 
\begin{equation}
    B=0,\quad \frac{1}{\Gamma(-\nu)} = 0 \implies \nu \in \mathbb{N}
\end{equation}
to get the correct behaviour. \\
This gives us the definition of the matching coefficient for $\phi_-$
\begin{equation}
     A_-\, D_n(t(x_1)) = \frac{1}{Q^{1/4}(x_1,s)},
\end{equation}
\begin{equation}
    \phi_-(x,s) =
    \begin{cases}
        \frac{1}{Q^{1/4}(x,s)}\exp{(i\int_{x}^{x_1} \sqrt{Q(x',s)}\;{\rm d}x')},\quad &x<x_1\\
        A_- D_n(t(x))= \frac{D_n(t(x))}{Q^{1/4}(x_1,s)D_n(t(x_1))}, \quad &x_1<x<x_2.
    \end{cases}
\end{equation}
Instead to match the solution from $+\infty$ (i.e. $\phi_+$) is sufficient to impose only
\begin{equation}
    B=0,
\end{equation}
and $\nu$ could be generic, hence leading to
\begin{equation}
     A_+\, D_{\nu}(t(x_2)) = \frac{1}{Q^{1/4}(x_2,s)}, \quad \nu \in \mathbb{C},
\end{equation}
\begin{equation}
    \phi_+(x,s) =
    \begin{cases}
        \frac{1}{Q^{1/4}(x,s)}\exp{(i\int_{x_2}^{x} \sqrt{Q(x',s)}\;{\rm d}x')},\quad &x>x_2\\
        A_+ D_{\nu}(t(x))= \frac{D_{\nu}(t(x))}{Q^{1/4}(x_2,s)D_{\nu}(t(x_2))}, \quad &x_1<x<x_2.
    \end{cases}
\end{equation}
Now there is a little caveat that should be addressed. From the equations above we have that 
\begin{equation}
    Q(x,s) = i \left( n+ \frac 12 \right) \sqrt{2Q''(x_0)} + \frac{Q''(x_0)}{2} (x-x_0)^2,
\end{equation}
which means that the inversion points $Q(s,x=0)$ are complex. 
Furthermore we need to impose the continuity in a point outside $x_1$ and $x_2$. 
A possible solution is to take $x_1$ and $x_2$ on the real axis where we have the correct oscillatory behaviour from the WKB solution.
This is reasonable because it is here that the parabolic cylinder $D$ function has the correct behaviour so it could be matched. The behaviour that we want to match, like in Ref. \cite{Schutz:1985km}, requires that the quadratic part is bigger than the imaginary part, i.e.
\begin{equation}
    \left( n+ \frac 12 \right) \sqrt{2Q''(x_0)} \ll \frac{Q''(x_0)}{2} (x-x_0)^2,
\end{equation}
but we also need to stay within the parabolic approximation range, because otherwise we end up in the asymptotic region which evolves like a free wave.

An estimate can be done by considering the third derivative and where this becomes important in the expansion, i.e. when
\begin{equation}
    \frac{1}{6} Q'''(x_0)(x-x_0)^3 \sim \frac{1}{2} Q''(x_0)(x-x_0)^2,
\end{equation}
which identifies the last point where this matching can be performed,
\begin{equation}
    x_{\rm max} \sim  \frac{3 Q''(x_0)}{Q'''(x_0)} + x_0.
\end{equation}
Hence we can choose a point in the range
\begin{equation}
    2 \epsilon\sqrt{ n+ \frac 12 } \;\frac{1}{(2Q''(x_0))^{1/4}}  < |x - x_0| <  \frac{3\, Q''(x_0)}{Q'''(x_0)} ,
\end{equation}
with $\epsilon \ll 1$. We define $x_1$ and $x_2$ as those points respecting this matching property. Now we go on with the matching starting from $-\infty$ and we match $\phi_-$ between the region $I_2$ and the region $I_3$, to get the full solution in the whole space. In $I_3$ we have
\begin{equation}
        \phi^{(3)}_-(s,x) = \frac{T}{Q^{1/4}(x,s)}\exp\left({i\int_{x_2}^{x} \sqrt{Q(x',s)}\;{\rm d}x'}\right),
\end{equation}
where $T$ is the transmission coefficient.
We need to match it to 
\begin{equation}
    \phi^{(2)}_-= \frac{1}{Q^{1/4}(x_1,s) \; D_{n}(t(x_1))}\; D_{n}(t(x)),
\end{equation}
at the point $x_2$. So we have an explicit form for the transmission coefficient
\begin{equation}
\label{t}
     \frac{T}{Q^{1/4}(x_2,s)} = \frac{D_{n}(t(x_2))}{Q^{1/4}(x_1,s) \; D_{n}(t(x_1))},
\end{equation}
which could also be rewritten as
\begin{equation}
    T = \frac{A_-}{A_+}.
\end{equation}
For $\phi_+$ the calculation is similar and the result is the same if $\nu =n$, as expected in the linear case.

\begin{figure}
    \centering
    \includegraphics[scale=0.6]{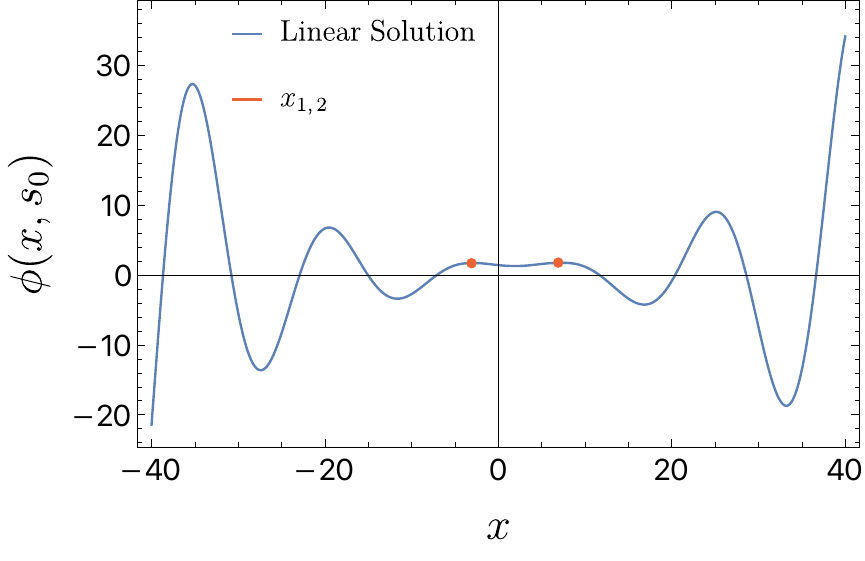}
    \caption{Real part of the WKB solution at linear level. Highlighted in red are the two inversion points. Notice that the function seems to diverge at $\pm \infty$, but when we reintroduce the time dependence and the causality conditions we end up with an asymptotically vanishing solution. The plot is in units $M=1$.}
    \label{fig:solution}
\end{figure}

\subsection{The Wronskian}
\label{sub_wronsk}
For our evaluation another ingredient is needed, i.e. the Wronskian, built from the functions $\phi_{\pm}$. As we've seen those are the solutions of the homogeneous equation,  but with only one of the two boundary condition satisfied. With those available we can find an approximate form for the Wronskian
\begin{equation}
    W(s) = \phi_- (x) \,\phi_+'(x) - \phi_+ (x) \,\phi_-'(x).
\end{equation}
This function is independent of the specific point picked, but we would like to extract as much information as possible by picking a specific easy point. In particular we would like to choose $x_0$, but to be able to do this we need to match the two WKB solutions to the central parabolic region.\\
We have already seen that the $\phi_-$ solution connected to the region within the inversion points is fully constrained,
\begin{equation}
\label{midphi-}
    \phi_-(x) = A_-\, D_n(t(x)), \quad x\in I_2,
\end{equation}
with
\begin{equation}
    A_-=\frac{1}{Q^{1/4}(x_1,s) \; D_{n}(t(x_1))}.
\end{equation}
For $\phi_+$ we have instead
\begin{equation}
\label{midphi+}
    \phi_+(x) = A_+\, D_{\nu}(t(x)), \quad x\in I_2,
\end{equation}
with
\begin{equation}
    A_+=\frac{1}{Q^{1/4}(x_2,s) \; D_{\nu}(t(x_2))}, \quad \nu \in \mathbb{C},
\end{equation}

because from the limit in Eq.  (\ref{pluslim}) is enough to put only $B=0$ to get the correct behaviour. We then have two different constants in front of the solutions $A_+$ and $A_-$.
The Wronskian is then 
\begin{align}
    W(s) = A_-\,A_+ \; \bigg{[} D_n(t) \left(\frac 12  t D_{\nu}(t) - D_{\nu+1}(t)  \right) \frac{\partial t}{\partial x} - \\ D_{\nu}(t) \left(\frac 12  t D_{n}(t) - D_{n+1}(t)  \right) \frac{\partial t}{\partial x}  \bigg{]} \Bigg|_{x_0},
\end{align}
which, remembering that $t(x_0) = 0$, gives
\begin{align}
    W(s) &=& \;A_+ A_-\; 2^{(n+\nu +1)/2} \,\pi \;(\partial t / \partial x) \;\;\times \nonumber\\
    &\times&\; \left(\frac{1}{\Gamma\left(\frac{1-n}{2}\right)\Gamma\left(\frac{-\nu}{2}\right)} - \frac{1}{\Gamma\left(\frac{1-\nu}{2}\right)\Gamma\left(\frac{-n}{2}\right)}\right)
\end{align}
which is zero only if $\nu \,= \,n$, i.e.
\begin{equation}
    i\; \frac{s^2 + V(x_0)}{\sqrt{2Q''(x_0)}} -\frac 12  = n.
\end{equation}
This is the form of the Wronskian we used to evaluate the non-linearities in Section \ref{sec_WBKnonlin}.\\
We could also expand for $\nu$ around $n$, to show that at linear order we get a zero around $s-s_0$. The Wronskian around the zero becomes
\begin{align}
    W(s) = \frac{ 2 \pi i\; 2^{n + 1/2} \,e^{i \pi/4}}{Q^{1/4}(x_1,s_0)Q^{1/4}(x_2,s_0) \; D_{n}(t(x_1)) D_{n}(t(x_2))}\; \times \nonumber\\
    \times \;\sqrt{\frac{\pi}{2}}  \frac{s_0 \Gamma(n)}{(2Q''(x_0))^{1/4} } \left(s - s_0\right) + \mathcal{O}\left((s - s_0)^2\right)
\end{align}
where we have defined $s_0$ as the solution for a specific $n$ and we evaluated $\nu=n$ in the $A_+$ coefficient.
Notice that the same Wronskian but at nonlinear level does not have a zero in $s = 2s_0$ and in general the $\nu$ dependence will lead to both a real and imaginary part.\\

\bibliography{Draft}

\end{document}